\documentclass{raa}            

\usepackage{graphicx,times}             
\usepackage{natbib} 
\begin{document}

   \title{On the distributions of pulsar glitch sizes and the inter-glitch time intervals.
}

   \volnopage{Vol.0 (200x) No.0, 000--000}      
   \setcounter{page}{1}          

   \author{I. O. Eya
      \inst{1,3}
   \and J. O. Urama
      \inst{2,3}
   \and A. E. Chukwude
      \inst{1,2,3}
   }

   \institute{Department of Science Laboratory Technology, University of Nigeria, Nsukka, Nigeria {\it innocent.eya@unn.edu.ng}\\
        \and
             Department of Physics and Astronomy, University of Nigeria, Nsukka, Nigeria\\
        \and
             Astronomy and Astrophysics Research Lab, University of Nigeria, Nsukka, Nigeria\\
   }
   \date{Received~~2017 XXXX XXXX; accepted~~XXXX}

\abstract{The glitch size, $ \Delta\nu/\nu $, inter-glitch time interval, $ t_{i} $ and frequency of glitches in pulsars are key parameters in discussing glitch phenomena.
In this paper, the glitch sizes and inter-glitch time intervals were statistically analysed in a sample of 168 pulsars with a total of 483 glitches. 
The glitches were broadly divided into two groups. Those with $ \Delta\nu/\nu < 10^{-7} $ are regarded as small size glitches, while those with $ \Delta\nu/\nu \geq 10^{-7} $ are regarded as relatively large size glitches.
In the ensemble of glitches, the distribution of $ \Delta\nu/\nu $ is seen to be bimodal as usual.
The distribution of inter-glitch time intervals is unimodal and the inter-glitch time intervals between small and large size glitches are not significantly different from each other. 
This observation shows that inter-glitch time intervals are size independent.
In addition, the distribution of the ratio $ \Delta\nu/\nu:t_{i} $ in both small and large size glitches has 
the same pattern. 
This observation suggests that a parameter which depends on time, which could be 
the spin-down rate of a pulsar plays a similar role in the processes that regulate both small and large size glitches.
Equally this could be an indication that a single physical mechanism, which could produce varying glitch sizes at similar time-intervals could be responsible for both classes of glitch sizes.
}

   \authorrunning{I. O. Eya, J. O. Urama \& A. E. Chukwude}            
   \titlerunning{On the distributions of pulsar glitch sizes and inter-glitch time intervals in pulsars.}  

   \maketitle

%
%

\section{Introduction}           
\label{sect:intro} 
Pulsars are believed to be spinning neutron stars \citep{b8a}, core remnants of supernovae events \citep{b4a,b4b}. 
They are very compact and
their central densities exceed the standard saturation nuclear density. This is a feature that makes them the second densest object observed in the universe. 
The internal temperatures of radio pulsar are thought to be below the neutron
 superfluid transition temperature, suggesting that their interiors contain superfluid neutrons \citep{b11a,ba5,b14}. 
The superfluid neutrons spin at a higher velocity than the observed velocity of the
solid crust, enabling it to act as a reservoir of angular momentum as we shall discuss latter.

The pulsar glitch manifests as a sudden increase (spin-up) in spin frequency ($ \Delta\nu $) of otherwise a steady spinning down pulsar, which is sometimes accompanied by a change in the magnitude of the spin-down rate ($ \Delta\dot{\nu} $) \citep{bcu,b7,b33}.
For many pulsars, one glitch is usually followed by a recovery phase 
during which the pulsar returns to a steady spin-down state on a wide range of time scale \citep{b32,b10}.  
These behaviours could be 
attributed to an internal process of a pulsar, which makes the pulsar glitch to be a feasible tool for studying the dynamic behaviour of super-dense matter (such as neutron superfluid) inside the star.

For nearly half a century since the first known pulsar glitch was discovered, the causes of pulsar glitches have been still
speculative despite numerous works put forward in explaining these events .
There could be external and internal origins for pulsar glitches,
but for the fact that early known glitches in rotation-powered pulsars were not accompanied by radiative changes, the causes of glitches were strongly attached to the internal mechanism of pulsars [for a recent review, see \cite{b46b}].
These causes can be broadly grouped into two depending on the mechanism involved, namely: the star-quake model and the angular momentum transfer model.
The star-quake model relies on a sudden reduction in the pulsar
 moment of inertia arising from the change in the oblateness of the star \citep[e.g.][]{b0,b01,b7fra}.
When the oblateness of a spherically rotating 
body is reduced instantaneously, such a body will spin up
 to compensate for the reduction in moment of inertia.
Such a model is convenient in explaining small-size glitches in young pulsars ($ \leq 10^{3} $ yr), such as the Crab pulsar \citep{b9a,b0}.  
However, for middle-aged pulsar ($ \approx 10^{4} $ yr) such as Vela pulsar, the star-quake model is not 
insufficient \citep{b01,b31}.
A more promising model for explaining pulsar glitches relies on the
exchange of angular momentum between star's components that are weakly coupled together \citep{b5,b3,b1}.
The components are the solid crust coupled to the core and the superfluid neutrons permeating into the inner-crust (crustal fluid) of the star. 
The crustal fluid viewed as angular momentum reservoir rotates via a quantized vertex array. 
These quantised vortices carry the circulation of the superfluid that it mimic the rotation of normal fluid.
The number of vortices contained in the superfluid determines its angular velocity.
These vortices have the ability to transfer their momentum to the other component if the right condition sets in.
Furthermore, these superfluid vortices are pinned in the lattice of the inner-crust \citep{b3}.
The pinned vortices enable the crustal fluid to partially decouple from the other components spinning at the same rate with the solid crust.
For as long as the vortices remain pinned, the superfluid velocity is conserved.
In this frame, as the crust spins down, the crustal fluid maintains its natural velocity.
The velocity difference between the two components (i.e. rotation lag) increases with time and could reach a point when the pinned vortices can no longer withstand the load on it due to the rotational lag.
As yet, unclear mechanism triggers a spontaneous unpinning of some (or the entire) vortices.
The unpinned vortices migrate outward transferring their momentum to the crust, the superfluid spins down and crust spins up (i.e. glitch) \citep{b3,b1}. 
Glitches occurring in this way should follow a canonical order in the sense that the accumulation and release of momentum is a function of pulsar rotational parameters (i.e. the spin frequency, $ \nu $ and the spin-down rate, $ \dot{\nu} $).
This implies that there should be a form of correlation between glitch sizes and inter-glitch time interval.
However, this logic is not observed in many glitch events.

On the other hand, recently, the relic from the early universe could also be a trigger of pulsar glitch \citep{bx}.
Such relic like ``strange nuggets", which mostly exist as dark matter could collide with pulsar to trigger a glitch, which resemble that of star-quake.
\cite{bx} has shown that the collision rate of pulsars with strange nuggets is consistent with the occurrence rate of small size glitches, though this model would not be able to explain glitches in pulsar with mixed glitch sizes.

Meanwhile, pulsar glitches are rare phenomenon. 
Since the first observed glitch in Vela and Crab pulsar (PSR J0835-4510 and J0534-2200) \citep{b12,b5b}, only few hundreds of such events have been recorded across the pulsar population with majority of them having few number of glitches per pulsar\footnote{for update, see Jodrell Bank Observatory (JBO) glitch database: http://www.jb.man.ac.uk/pulsar/glitches.html.}.
This low glitch rate, which characterizes the few multiple glitching pulsars, has hindered comprehensive statistical study of pulsar glitches. 
The pioneer glitching pulsars (i.e. Vela and Crab pulsars) have by far been the best studied.
Thus, they have become a reference point in discussing pulsar glitch events. 
This is partially because they were first to be identified with such event, and mainly due to the characteristic size and frequency of the glitches they exhibit.

Glitch sizes are in the range of $10^{-11} \leq \Delta\nu/\nu \leq 10^{-5}$ \citep{b7}, while the inter-glitch time intervals are in the range of 20 d $\leq t_{i} \leq  10^{4}$ d.
Most of the Vela pulsar glitches occupy the right end of the size distributions (with $ \Delta\nu/\nu \approx 10^{-6}$), while most of the Crab pulsar glitches occupy the left end (with $ \Delta\nu/\nu \approx 10^{-9}$).
Across the population of glitching pulsar, most pulsars with multiple glitches present a broad range of glitch sizes.
This has made it extremely difficult to use a single mechanism to explain all features of pulsar glitches across the population.
Predicting glitch sizes from inter-glitch time intervals have been quite challenging, as most events appear to be random in size.
The cumulative glitch sizes in Vela pulsar is quite linear with time \citep{b19c,b8,b9,b7c,b7e}, but this trend is not exhibited by glitches in Crab and many other pulsars.
However, evidence of linear dependence of cumulative glitch size with time exists in two other pulsars, namely, PSRs J0357-6910  \citep{b43a,b7e,bf,ba} and J1420-6048  \citep{b7e}.

Other statistical analyses of glitch parameters in ensemble of pulsars have also turned out interesting results. Such as: correlation of glitch activity with pulsar spin-down rate \citep{b51,buo,b7} and the spin-down luminosity \citep{b7fue}, glitch size following a power law with respect to energy released in the event \citep{bmo} and young pulsars having high glitch rate than old pulsars \citep{b7}. 
\cite{b11} analyzed glitch sizes and waiting times in individual pulsars using the cumulative distribution function (CDF) to show that glitch events could be reproduced in a glitch mechanism that involves avalanches of vortices. In the model, the basic assumption is that the neutron superfluid vortices are in a self-organized criticality (SOC) state.  As such, the glitch size distribution is seen to be power law, while the waiting times distribution is exponential. 
Following \cite{b11}, \cite{bonu} presented a similar result using a dataset of microglitches recorded Hartebeesthoek radio telescope. 
Also, using the CDF plot, \cite{b7e} demonstrated that the cumulative spin-up sizes in 12 pulsars follows a normal distribution. These pulsars are mostly pulsars in which the dispersions in their glitch sizes distribution are narrow.

In addition, the histograms of glitch size distribution have consistently been bimodal, \citep[e.g.][]{b31a,b7,b33,b7c,b7e} there-by signaling a dual glitch mechanism.
Nonetheless, the relatively large size glitches are mostly from few middle age pulsars. 
In that, it was suggested that this bimodal nature of the distribution could result from two different classes of pulsars or via a glitch mechanism, which advances as the pulsars grow old. 
However, this bimodal distribution of glitch sizes, which have been widely reported, is in contrast to that of the spin-up sizes ($ \Delta\nu $), which tends towards multimodal distribution \citep{b7}.  
With a mixture of Gaussians fitted in the histogram of spin-up sizes, \cite{b7fue} also reported on it. 
Recently, from the estimation of glitch sizes and waiting times distributions, there is no persuasive proof for such multimodal distribution \citep{bhow}. 
Using the distribution of fraction of neutron star components participating in glitch, \cite{b7e} showed that the missing glitches that caused the dip in the histogram of glitch size distribution is well accommodated within the range feasible with the angular momentum transfer models. 
As such, the bimodal in the distribution of glitch sizes may not be connected to glitch mechanism.
These contrasting views may linger longer than expected due to paucity of glitch events, though it is expected that a clearer picture will emerge any time glitch data improves significantly especially in individual pulsars. 
In all these analyses, efforts have not been made towards studying the distribution of inter-glitch time interval in ensemble of pulsars or in analyzing the glitch data corresponding to the peaks of the histograms of glitch size distribution. 
Such a study could present a clear picture to ascertain whether inter-glitch time intervals are size dependent or whether the peaks in the histograms of glitch size distributions could be linked to glitch mechanisms or not.
In this paper, we shifted from the previous approach by making the distributions of time intervals between glitches and analyse the
inter-glitch time intervals corresponding to small ($ \Delta\nu/\nu < 10^{-7} $) and relatively large ($ \Delta\nu/\nu \geq 10^{-7} $) size glitches
using an ensemble of 483 glitches in 168 pulsars.
Equally, we investigate the distribution pattern of $ \frac{\Delta\nu/\nu}{t_{i}} $ in both small and large sized glitches, where $ t_{i} $ is the inter-glitch time interval preceding a given glitch.

\section{Data Analyses and Results}
\label{sect:Obs}
The glitches for this analysis were taken from the Jodrell Bank Observatory glitch catalogue\footnote{http://www.jb.man.ac.uk/pulsar/glitches.html,}.
The catalogue contains 483 glitches in 168 pulsars, which are known as at the time of this analysis.
The glitch sizes are in the range of $ 10^{-11} \leq \Delta\nu/\nu < 10^{-4}$ while the corresponding change in the spin-rate is in the range of $10^{-6} \leq |\Delta\dot{\nu}/\dot{\nu}| \leq 10^{0} $. 

We assume that the observed sudden increases in pulsar spin frequency during glitches result from transfer of angular momentum stored in momentum reservoir over a glitch-free period preceding the glitch (hereafter called the inter-glitch time interval, $t_{i}$).
With this, we evaluate the inter-glitch time interval for n$^{th}$ glitch as $ t_{i}^{th} = t_{n} - t_{n-1} $, where $ t_{n} $ is the Modified Julian date for n$^{th}$ glitch and $ t_{n-1} $ is the Modified Julian date for (n-1)$^{th} $ glitch.
The histogram of the distributions of glitch sizes, $ \Delta\nu/\nu $, and inter-glitch time intervals, $ t_{i}$, are shown in Fig. 1.
Both distributions are continuous: the distribution of $ \Delta\nu/\nu $ (top panel) is bimodal as has been widely reported \citep[e.g.][]{b31a,b7,b33,b7c,b7e} with peaks at $ \Delta\nu/\nu \approx 2.14 \times 10^{-9} $ and $ \Delta\nu/\nu \approx 1.17 \times 10^{-6} $. 
On the other hand, the distribution of $ t_{i}$ is unimodal with peak at $ t_{i} \approx 2.96 \times 10^{3}$ d.
The histogram of $ t_{i} $ approximates that of a normal distribution as suggested by the fit on it. 
The mean and median $ t_{i} $ are coincident in the bin before the modal bin.
From the histograms (Fig. 1) there is no sign of larger $ t_{i} $ corresponding to larger $ \Delta\nu/\nu $ or vice verse.

\begin{figure}
\centering
\includegraphics[scale=0.6]{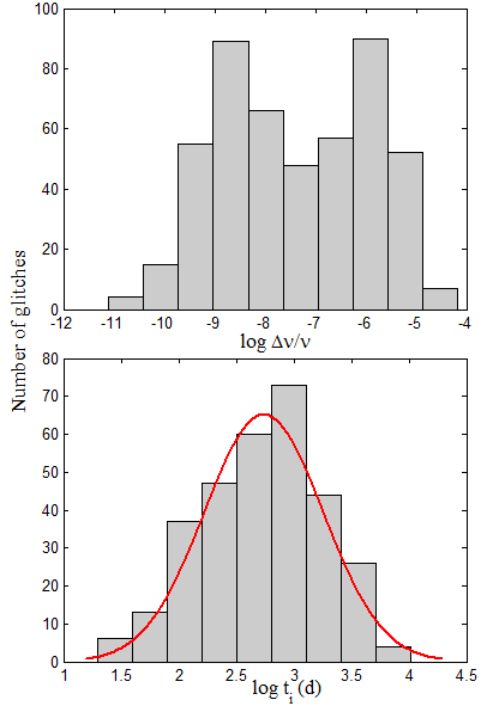}
\caption{Top panel: distribution of glitch sizes ($ \Delta\nu/\nu $). Bottom panel: distribution of inter-glitch time intervals ($ t_{i} $).}
\end{figure}

In order to estimate a possible demarcation between relatively large and small size glitches, we fitted twin gaussians on the glitch size distribution. These gaussians are fitted as such that their modes are within the modal peak of each side of the glitch size distribution. 
Then the measure of central tendency of the gaussians, the portion of the glitch size distribution they demarcate and other shape parameter of  the portion is evaluated. 
This helps to determine the potion of the glitch size distribution, which has a similar characteristic with the gaussian.
In addition, our interest is on the fits in which the difference between the mode of the portion it demarcate and the mode of the gaussian is not more than $\pm \frac{1}{2}w$, where $w$ is the bin-width of the histogram of the glitch size distribution\footnote{This will ensure that each portion do not deviate much from normal distribution. The bin-with is 0.69}.  
The result is summarised in Table 1 and
the distributions with the best fits are shown in Fig. 2.
From Fig. 2, it is readily observed that the fit in bottom panel gives a better demarcating line between large size glitches and small ones than other panels. 
The mean of the gaussians coincide with the mean of the portions they demarcate.
The standard deviation of the gaussians and that of the portion they demarcate are within the same range.
Also the kurtosis\footnote{Note: the kurtosis of normal distribution/gaussian shape is 3} of the portions is closer to that of a normal distribution than others. 
The twin gaussians intercepts at approximately $ \Delta\nu/\nu = 10^{-7}$. 
Consequently, glitches with $ \Delta\nu/\nu < 10^{-7}$ are regarded as small-size glitches (SSG)  and large-size glitches (LSG) are taken as those with $ \Delta\nu/\nu \geq 10^{-7} $.

\begin{table}
\caption{
A measure of the central tendency of the two peaks of the glitch size distribution with respect to the guassian fits in Figure 2.}
   \begin{center}  
  \begin{tabular}{@{}|cccc|@{}}
  \hline Statistic & Top panel &Middel panel&Bottom panel\\
 &LP \, $ $ $  $ $ $ \,RP & LP \,$ $ $  $ $ $\, RP &LP \, $ $ $  $ $ $ \,RP\\
     		\hline
$ \overline{\mu}_{p} $ &	-9.40 \,	-6.01 &	-9.06 \,-5.96	 &-8.54	 \,-5.93	\\
$ \overline{\mu}_{g} $ &	-8.79 \,	-6.21&	-8.54 \, -6.23	&-8.54 \,-5.93\\

$\mu_{p}$   &	-9.22\,	-5.73  &-8.96\,	-5.95	&-8.55 \,	-5.93\\
$\mu_{g}$	 &	-8.79 \,	-6.21&	-8.54	 \,	-6.23 &	 -8.54 \,-5.93\\

$\rho_{p}$   &	-8.67\,  -5.93  &-8.67\, -5.93	&-8.67 \,-5.93\\
$\rho_{g}$	 &	-8.79 \,	-6.21 &	-8.54	 \,	-6.23 &	 -8.54 \,-5.93\\

$ \sigma_{p} $       &	1.08 \, 1.10 & 1.27  \,	1.12 & 1.32 \, 0.55 \\
$ \sigma_{g} $       &	1.20 \,  1.36 & 1.43	 \,1.39 &	1.40 \,0.60\\

 $ ku $              &	2.10 \,2.30       & 2.60 \,2.25	& 2.62\, 2.87\\
\hline
\end{tabular}
\end{center}
\fontsize {8}{1.8}\selectfont
Note: LP (RP) denotes the portion of the glitch size distribution of which the Left Peak (Right Peak) is assumed to be its mode.
$\overline{\mu}_{p/g}$ denotes the mean of the portion/gaussian, $\mu_{p/g}$ the median of the portion/gaussian, $\rho_{p/g}$ the mode of the portion/gaussian, $ \sigma_{p/g} $ the standard deviation of the portion/gaussian and $ ku $ denote the kurtosis of the portion demarcated by the gaussian.
\end{table}

In order to investigate whether the 
inter-glitch time intervals\footnote{i.e. the glitch free-interval preceding that glitch, independent of the size of the bracketing glitch} are
actually size independent as envisage in Fig. 1, we 
made two classes of inter-glitch time intervals with respect to the two class of glitches and explore the two dimensional Komogorov-Smirnov (K-S) test.
Those one corresponding to SSGs we denote with $ _{S}t_{i} $, and those corresponding to LSGs denoted with $ _{L}t_{i} $. 
The K-S test enables us to determine the similarity in the distributions the classes of the inter-glitch time intervals (i.e. $ _{S}t_{i}$ and $ _{L}t_{i} $.)

In K-S test, a null hypothesis (i.e. h = 0) is that in a given dataset, $ x_{1} $ and $ x_{2} $ described by CDFs $ F_{1} (x) $ and $ F_{2} (x) $ are drawn from the same continuous distribution $x$.
An alternative hypothesis (i.e. h = 1) is that the two data set are drawn from different continuous distributions.
The K-S statistics D is the maximum distance between $ F_{1} (x) $ and $ F_{2} (x) $.
The probability that the result is by chance is ascertained by the magnitude of a P-value. 
The null hypothesis is rejected if the P-value is small, but a small P-value is a strong evidence for accepting the alternative hypothesis\footnote{Note: the values of D and P are in the range of $0 - 1$}.  
In this paper, the K-S tests are performed at 5\% significant level.
If the null hypothesis is true, 
it is an indication that the two distributions are drawn from a common continuous distribution, implying that inter-glitch time intervals are size independent.
In addition, it could be an indication that a parameter that involved time is scale invariant and plays the same role in the process that cumulated in the two classes of glitch sizes.

\begin{figure}
\centering
\includegraphics[scale=0.5]{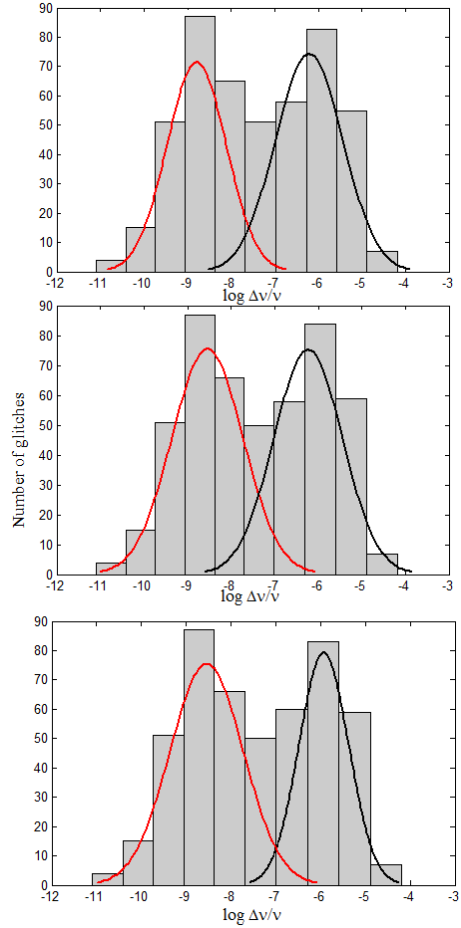}
\caption{Distribution of $ \Delta\nu/\nu $ with the twin gaussian fits. The modes of the gaussians are chosen to be within the modal bins of the distribution.}
\end{figure}

The results obtained from the application of K-S test to the distributions of the two classes of the inter-glitch time intervals are as follows: 
h = 0, D = 0.07 and P = 0.88.
The 
CDF is shown in Fig. 3.
The outcome of this preliminary test indicates that  
there is no significant difference between the distributions of $ _{S}t_{i} $ and $ _{L}t_{i} $. 
The maximum distance between the two curves is very small (Fig. 3) and a large P-value, enabling us to accept the null hypothesis. 
This is an indication that, on average, the inter-glitch time intervals are size independent with respect to glitch size.
Furthermore, a parameter, which is strongly tied to time, is scale invariant and plays a similar role in the process that produce SSGs and LSGs.

\begin{figure}
\centering
\includegraphics[scale=0.7]{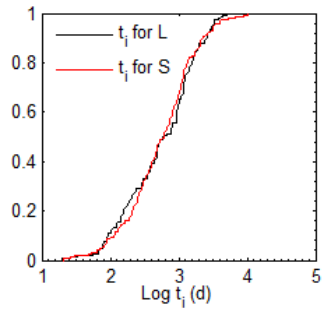}
\caption{CDFs of inter-glitch time intervals ($ t_{i} $) corresponding to LSGs and SSGs.}
\end{figure}

Next, following \cite{b51} who defines the quantity, $ \frac{\Sigma(\Delta\nu/\nu)}{t_{g}} $, as the `glitch activity', where $ t_{g} $ is total time for the cumulative glitch size, $ \Sigma(\Delta\nu/\nu) $, we define the quantity, $ \frac{\Delta\nu/\nu}{t_{i}} $, as the `unit activity'.
As the distribution of $ t_{i} $ in both class of glitch size is not significantly different from each other, the distribution pattern of $ \frac{\Delta\nu/\nu}{t_{i}} $ in both classes should be similar if both class of glitch sizes originate from a single process, which is time invariant with respect to size. 
To investigate this, the distribution of unit activities in LSG and SSG were subjected to normality test\footnote{A normality test determines whether in a given dataset, if the elements in the dataset are similarly sized and symmetrical about a mean value.}.

In this analysis, we used Lilliefors test at 5\% significant level for the normality test.
Unlike the one dimensional K-S test, the Lilliefors test does not require to predetermine the type of normal distribution, instead the type of normal distribution is determined from the dataset \citep{b15,b16}.
A null hypothesis in Lilliefors test (i.e. h = 0), is that the elements in the dataset are drawn from a normal distributed population or else, otherwise.

On application of the normality test on the unit activities, the results indicate that the distributions approximate normal distribution in both classes.
The P-values are 0.90 and 0.78 for LSG and SSG respectively.
The CDF plots are shown in Fig. 4. 
The fit on the CDF plots is based on normality test result to indicate how the data fits the distribution.
The goodness of the fits, were quantified numerically using the two dimensional K-S test.
The test yield P/D-values of 0.99/0.01 and 0.80/0.04, for LSG and SSG respectively enabling us to accept the null hypothesis.
The outcome of this test signifies that a parameter, which depends on time, plausible the spin-down rate of pulsar plays a similar role in the processes that regulate both small and large size glitch. 
This is because dividing the glitch sizes with their corresponding inter-glitch time intervals produces the same effect on the distribution pattern of both large and small size glitches.
Equally, it reaffirms that the inter-glitch time intervals are size independent.

\begin{figure}
\centering
\includegraphics[scale=0.7]{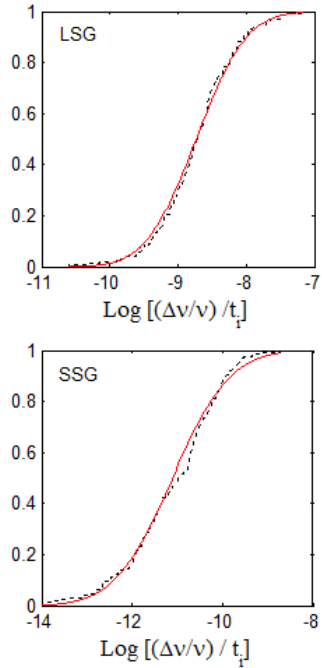}
\caption{Distribution of of unit activity: Top panel corresponds LSG, while bottom panel corresponds to SSG. 
Dotted lines indicates the CDF of the unit activity, while solid lines indicates the CDF of ideal normal distribution that has the same mean and standard distribution as the distribution of the unit activity.}
\end{figure}

\section{Discussion}
\label{sect:data}
Pulsar glitch sizes and inter-glitch time intervals are key elements in discussing glitch events. 
From the glitch data many pulsars do exhibit a mixture of glitch sizes, though majority of them have few glitches. 
SSGs are usually attributed to mechanism involving neutron star-quake, while LSGs are attributed to mechanism involving transfer of angular momentum.
The dip in the distribution of glitch sizes (Fig. 1) has remained one of the strongest evidences in favour of dual glitch mechanisms \citep{b33}.
However, this idea has an implication as it could suggest that more than one glitch mechanism operate in a pulsar with a mixture of glitch sizes.
If one regards the SSGs as events produced by a physical mechanism other than the angular momentum transfer, such as the star-quake, it could be trivial to state that two different independent physical processes (i.e. star-quake and angular momentum transfer processes) within a closed system have the similar time intervals between them.

Meanwhile, in the frame of angular momentum transfer, a recent analysis has shown that the fraction of neutron star components participating in glitch could account for the current range of glitch sizes, including the missing glitches that caused the dip in the distribution of glitch sizes \citep{b7e}. 
As such, mechanisms involving transfer of angular momentum could readily account for all sizes of pulsar glitches.
The glitch size has been shown to depend on three main factors: the number of vortices involved, the location of the vortices and the distance the vortices travel before re-pinning \citep{b31wa}.
Conventionally, the vortices are located in the inner-crust superfluid. 
The maximum distance the unpinned vortices could migrate before re-pinning is the thickness of the inner-crust, (assuming a spherical symmetric inner-crust, and the vortices migrate radial outward after unpinning).
In this regards, the distribution of the inter-glitch time intervals preceding LSG and SSG being the same could be feasible.
This is because the electromagnetic braking torque on the stellar crust, which induces the differential rotation that culminate in glitches is not primarily connected to the vortices. 
If its magnitude is fairly stable at much longer time compare to the inter-glitch time intervals, it could lead to similar inter-glitch time interval for both large and small size glitches.
The difference between LSG and SSG could be related to number of vortices involved or the distance the vortices migrate before repining. 
LSG could result from a situation in which almost the entire vortices were unpinned; they migrate the maximum distance before repining. 
On the other hand, SSG could result from a situation in which a bit of the vortices were unpinned at each event. 
The vortices move less distance before repining as their motion could be impeded by other pinned vortices. In such a scenario, a given pulsar could have a mixture of glitch sizes depending on how the vortex unpinning trigger mechanism operates in it.

Dividing the glitch sizes in each class with their corresponding inter-glitch time interval just normalized the two distributions with respect to time (Fig. 4).  
This observation reaffirms the similarity of inter-glitch time intervals in two classes of glitches. 
If one is to present this picture in a model induced 
by a similar physical process, a mechanism that could produce varying glitch sizes of similar time intervals should be sort for.
In addition, considering radiative changes associated with glitches in magnetars, and a few high magnetic field pulsars \citep{b6,bwje,bdk,bka} and recently in radio pulsar \citep{bkou}, such a mechanism should take into account magnetospheric activity.
Magnetar glitches are part of the LSG distribution. 
Ordinarily with respect to rotation-powered pulsar of similar characteristic age, glitches in magnetars should be smaller compare to what is observed. Magnetars are mostly young pulsars powered by the decay of their conspicuous magnetic field rather than the lost of their rotational energy \citep{bg,bzhu,bliu}, as such, there could be a contribution of magnetospheric activity to their glitch sizes \citep{bgao}.

Magnetospheric activity could come into play in glitches due to external torque on the pulsar coupled to it.
The gross torque on a pulsar is of two sources, namely, the external electromagnetic braking torque and the internal braking torque coming from the neutron superfluid vortex creeping \citep{byuan}. 
The external torque on the whole star causes small or micro-size glitch/timing noise and noticeable radiative changes, while the internal braking torque causes large size glitches, small size glitches and even timing noise \citep{byuan}.
As such, there could be external and internal origins for pulsar glitches depending on the contributions of torques involved, or even a combined action leading to large size glitches accompanied with noticeable radiative changes.

Moreover, the narrowing of pulse profile and abnormal evolution of spin-down rate ($|\dot{\nu}|$)  associated with glitch activity in rotation-powered pulsar, PSR J2037+3621 (B2035+36) could not be explained by standard glitch models \citep{bkou}. 
This is also a pointer that in some glitches, factors due to external mechanism could be significant.
In order to explain this anomaly, observed in PSR J2037+3621 \cite{bkou} assume that there could be a change in external braking torque accompanied with the glitch activity.
As the dipole magnetic field strength increases, the pulsar's braking index decreases due to an increasing braking torque and vice versa \citep{bga,bgb}.
A change in braking torque affects the magnetic field structure, which in turn affect the inclination angle\footnote{the angle between magnetic axis and rotational axis}. 
If this effect results from fluctuations in particle density outflow in the magnetosphere \citep{bkt}, fluctuation in the inclination angle shall lead to changes in magnetic field structure leading to radiative changes accompanying a glitch.

In conclusion, the distribution of the inter-glitch time intervals to have a single mode and that of $ _{L}t_{i} $ and $ _{S}t_{i} $ to be the same, is a clear indication that a global parameter involving time, plausibly the pulsar spin-down rate $ (\dot{\nu} $), plays the same role in both classes of glitches.
Parameters that strongly regulate the glitch size is not strongly tied to the time parameter, rather, 
it has much to do with the glitch mechanism. 
Therefore, in searching for an explanation for pulsar glitch events, a hybrid mechanism that has much to do with neutron star interior, taking account of radiative change as well as having the ability to produce different sizes of glitch at a similar time intervals, should be sort for.

\label{lastpage}

\end{document}